\documentstyle[12pt]{article}                                                 
\begin{document}     
\begin{center}{\Large\bf The First Six Days of the Universe}\end{center}
\begin{center}{Moshe Carmeli}\end{center}
\begin{center}{Department of Physics, Ben Gurion University, Beer Sheva 84105, 
Israel}\end{center}
\begin{center}{Email: carmelim@bgumail.bgu.ac.il}\end{center}
\begin{abstract}
The early stage of the Universe is discussed and the time lenghts of its first 
six days, as well as its age, are given. There seems to be no contradiction 
with the Bible's ascertain that the Universe was created in six days.
\end{abstract}
\newpage
Scientists have in recent years adopted the picture that the Universe was 
created in a single event of plasma explosion, called the Big Bang. This 
approach remarkably conforms with the Bible's description of the creation of
the Universe.
However, there are still doubts about the meaning, mentioned in the Bible, that
the Universe was created in six days. We actually know from the study of 
anthropology and cosmology that any development of the kind mentioned in the
Bible takes millions or billions of years. 

We show in the following that the 
viewpoint of the Bible is actually compatible with the theory of cosmology --
the days of our life now are not equal to the days at the time of the creation
of the Universe. In this note we calculate the lengths of days of the early
Universe, day by day, from the first day on up to our present time. We find 
that the first day actually lasted the Hubble time in the limit of zero 
gravity. If we denote the Hubble time in the zero-gravity limit by $\tau$ 
which equals 11.5 billion years and $T_n$ denotes the length
of the $n$-th day in units of times of the early Universe, then we have a very
simple relation 
$$T_n=\frac{\tau}{2n-1}.\eqno(1)$$
Hence we obtain for the first six days the following lengths of time:
$$T_1=\tau,\hspace{5mm}T_2=\frac{\tau}{3},\hspace{5mm}T_3=\frac{\tau}{5},
\hspace{5mm}T_4=\frac{\tau}{7},\hspace{5mm}T_5=\frac{\tau}{9},\hspace{5mm}
T_6=\frac{\tau}{11}.\eqno(2)$$

It also follows that the accumulation of time from the first day to the second,
third, fourth, ..., up to now is just exactly the Hubble time. The Hubble time
in the limit of zero gravity is the maximum time allowed in nature. Using 
Cosmological Special Relativity [1-4], the calculation is very simple. We 
assume that the Big Bang time with respect to us now was $t_0=\tau$, the time 
of the first day after that was $t_1$, the time of 
the second day was $t_2$, and so on up to the sixth day $t_6$. In this way the
time scale is progressing in units of one day (24 hours) in our units of 
present time. The time difference between $t_0$ and $t_1$, denoted by $T_1$, 
is the time as measured at the early Universe and is by no means equal to one
day of our time. In this way we denote the times elapsed from the Big Bang to 
the end of the first day $t_1$ by $T_1$, between the first day $t_1$ and the 
second day 
$t_2$ by $T_2$ and so on for the time from the fifth day $t_5$ to the sixth 
day $t_6$ by $T_6$. According to the rule of the addition of cosmic times one has
$$t_6+1(day)=\frac{t_6+T_6}{1+t_6T_6/\tau^2}.\eqno(3)$$
A straightforward calculation then shows that
$$T_6=\frac{\tau^2}{\tau^2-\left(\tau-6\right)\left(\tau-5\right)}=
\frac{\tau^2}{11\tau-30}.\eqno(4)$$

In general one finds that
$$T_n=\frac{\tau^2}{\tau^2-\left(\tau-n\right)\left(\tau-n+1\right)},\eqno(5)$$ 
or
$$T_n=\frac{\tau}{n+\left(n-1\right)-n\left(n-1\right)/\tau}.\eqno(6)$$
As is seen from the last formula one can neglect the second term in the 
denominator in the first approximation and we get the simple Eq. (1).

From the above one reaches the conclusion that the age of the Universe
exactly equals the Hubble time in vacuum $\tau$, i.e. 
11.5 billion years, and it is a universal constant. This means that the age 
of the Universe tomorrow will be the same as it was yesterday or today. 
But this might not go along with our intuition since we usually deal with short
periods of times in our daily life, and the
unexperienced person will reject such a conclusion. Physics,
however, deals with measurements. In fact we have exactly a similar situation
with respect to the speed of light $c$. When measured in vacuum, it is 300
thousands kilometers per second. If the person doing the measurement tries to 
decrease or increase this number by moving with a very high speed in the 
direction or against the direction of the propagation of light, he will find
that this is impossible and he will measure the same number as before. The
measurement instruments adjust themselves is such a way that the final result
remains the same. In this sense the speed of light in vacuum $c$ and the 
Hubble time in vacuum $\tau$ behave the same way and are both universal 
constants.

The similarity of the behavior of velocities of objects and those of cosmic 
times can also be demonstrated as follows. Suppose a rocket moves with  the
speed $V_1$ with respect to an observer on the Earth.
We would like to increase that speed to $V_2$ as measured by the observer on 
the Earth. In order to achieve this, the
rocket has to increase its speed not by the difference $V_2-V_1$, but by 
$$\Delta V=\frac{V_2-V_1}{1-V_1V_2/c^2}.\eqno(7)$$
As can easily be seen $\Delta V$ is much larger than $V_2-V_1$ for velocities
$V_1$ and $V_2$ close to that of light $c$.
This result follows from the rule for the addition of 
velocities,
$$V_{1+2}=\frac{V_1+V_2}{1+V_1V_2/c^2},\eqno(8)$$ 
a consequence of Einstein's famous Special Relativity Theory [5].
In cosmology, we have the analogous formula
$$T_{1+2}=\frac{T_1+T_2}{1+T_1T_2/\tau^2}\eqno(9)$$
for the cosmic times.

In conclusion, the lengths of the first six days were enough to accomodate the
activities of the creation mentioned in the Bible. Furthermore, since at that
time there were no other reference systems (like the present-day one) to 
compare with, one concludes that the ascertain of the Bible about the six-day 
creation of the Universe is scientifically valid. 
 
\end{document}